\documentstyle[aps,prb]{revtex}
\begin{document}
\draft
\title{Spin-orbit interaction in the quantum dot}
\author {Lucjan Jacak}
\address{Institute of Physics, Technical University of Wroc\l aw,
         Wybrze\.ze Wyspia\'nskiego 27, 50-370 Wroc\l aw, Poland}
\author {Jurij Krasnyj}
\address{Institute of Physics, University of Odessa,
         Petra Velikogo 2, Odessa 270100, Ukraina}
\author {Arkadiusz W\'ojs}
\address{Institute of Physics, Technical University of Wroc\l aw,
         Wybrze\.ze Wyspia\'nskiego 27, 50-370 Wroc\l aw, Poland}
\maketitle

\begin{abstract}
The electronic states of a parabolic quantum dot in a magnetic
field are studied with the inclusion of the spin-orbit interaction.
The analitycal formulae for the ground state energy of the
interacting system are derived.
The spin-orbit interaction is shown to introduce new features
to the far infra-red absorption spectrum, where it leads to the
splitting of the two principal modes.
The results are compared with the charging experiments
by Ashoori {\sl et al.} and the far infra-red absorption
measurements by Demel {\sl et al}.
\end{abstract}

\section{Introduction}

Recent progress in the semiconductor nanostructure technology has
allowed for creation of the quasi-zero-dimensional (0D) systems
known as quantum dots \cite{chak1}.
These structures can be obtained e.g. by applying spatially modulated
electric field to the 2D electron gas \cite{heitmann,alsmeier}, or by
embedding a small piece of one semiconductor into another, with higher
conduction band energy (this can be achieved e.g. through etching
\cite{demel,meurer,ashoori,bayer}, interdiffusion \cite{brunner} or
self-organized growth \cite{petroff,raymond}).
The resulting lateral potential is often very well approximated
by the isotropic parabola \cite{kumar,arek1,arek2}.
The confinement in all three spatial directions leads to a discrete
energy spectrum of the system with characteristic excitation energies
of the order of few meV, which can be studied using the spectroscopy
methods \cite{heitmann,alsmeier,demel,meurer}.

Due to the so called generalized Kohn theorem \cite{kohn,chak2} for
the parabolic confinement the electron-electron Coulomb interaction
does not affect the resonance energy spectrum under the far infra-red
(FIR) radiation.
The FIR resonant spectrum of the correlated many-electron dot is
therefore identical to that of a single electron and consists of
two degenerate modes.
The degeneracy can be removed by a perpendicular magnetic field
under which the two modes evolve into the inter-Landau-level
(cyclotron) and intra-Landau-level frequencies.
However, a number of experiments seem to show a slight dependence
of the two resonance energies on the number of electrons in the dot
\cite{alsmeier,demel,meurer}.
Moreover, a small splitting of the two principal modes is observed
\cite{demel} what reveals their additional sub-level structure.

The theoretical investigations of the ground state of quantum dots
containing 2--15 electrons were reported e.g. in Refs.
\onlinecite{chak2,pawel1,pawel2,arek3,darnhofer,juanjo}.
In this paper we calculate analitically the ground state energy of
a system of a larger number of confined electrons ($\sim$15--100)
within a Hartree-Fock approximation.
The spin-orbit interaction is included here in the manner analogous
to that used for many-electron atoms and not via the {\em bulk}
band-structure parameters, since the considered system is strongly
localized (diameter $\sim$20 lattice constants \cite{demel}) and
the small piece of crystal, to which the motion of electrons is
limited, cannot be treated as an infinite periodical lattice.
Instead, we propose to extract the spin-orbit coupling from the
band-structure description and include it later on the level of
the Bloch envelope wavefunctions by adding the appropriate term
to the hamiltonian.
As demonstrated by Darnhofer and R\"ossler \cite{darnhofer} for
2 electrons, inclusion of the spin-orbit interaction through the
band-structure parameters in the InSb dot leads to the similar
effects on the electronic structure as obtained here for the GaAs
dots (spin-orbit coupling in bulk InSb significantly exceeds that
in bulk GaAs).

The results obtained within proposed here framework seem to explain
a number of experimentally observed effects, like an appearance of
the higher modes in FIR absorption and their anti-crossing in a magnetic
field \cite{demel}, or the characteristic bumps in the magnetic-field
dependence of the ground state energy \cite{ashoori}.
Let us also underline that we managed to fit very well the characteristic
magnetic field at which the anti-crossing in FIR spectrum and the bumps
in energy occur.
The fact that this was impossible without including the spin-orbit
interaction (see the numerical results and discussion by Palacios
{\sl et al.} \cite{juanjo} for up to 15 electrons) seems to prove
the importance of this effect.

\section{Model}

The interaction between the spin and orbital angular momentum:
{\boldmath$\sigma$} and ${\bf l}$, of an electron confined in a quasi
two-dimensional quantum dot is included in the way analogous to that
used in many-electron atoms, i.e. via the single-particle potential:
\begin{equation}
  V_{LS}=\alpha{\bf l}\cdot\mbox{\boldmath$\sigma$}.
\end{equation}
The coupling constant $\alpha$ is connected with the average
self-consistent field $<\delta U>$ acting on the electron,
via the relation:
\begin{equation}
   \alpha=\beta<\delta U>.
\label{soc}
\end{equation}
For a $Z$-electron atom the dimensionless parameter $\beta$ is:
\begin{equation}
   \beta=\left({Ze^2\over\hbar c}\right)^2
   \approx\left({Z\over137}\right)^2.
\end{equation}
In the case of a quantum dot $\beta$ will be treated as a fitting
parameter, while the magnitude of the field $<\delta U>$ will
be estimated according to the electronic structure of the dot.

Let us write the total hamiltonian of a system of many electrons,
in the effective mass approximation, including the kinetic energy
in the perpendicular magnetic field $B$, parabolic confinement of
the characteristic frequency $\omega_0$, spin-orbit coupling as above,
the Zeeman splitting for the effective $g$-factor and the
electron-electron Coulomb interaction controlled by the dielectric
constant $\epsilon$:
\begin{eqnarray}
   {\cal H}&=&\sum_i\left[
   {1\over2m^*}\left({\bf p}_i+{e\over c}{\bf A}_i\right)^2
   +{1\over2}m^*\omega_0^2 {\bf r}_i^2
   +\alpha l_i\sigma_i - g\mu_B\sigma_iB \right]
   +{1\over2}\sum_{i\ne j}{e^2\over\epsilon|{\bf r}_i-{\bf r}_j|}
   \nonumber\\
   &\equiv&\sum_i(H_B)_i
   +{1\over2}\sum_{i\ne j}{e^2\over\epsilon|{\bf r}_i-{\bf r}_j|}.
\label{ham}
\end{eqnarray}
In the above $m^*$ is the effective mass, $\bf r$ are the positions,
${\bf p}=-i\hbar\nabla$ are the momenta, ${\bf A}={1\over2}B(y,-x,0)$
are the vector potentials in the symmetric gauge, $l$ and $\sigma$ are
the projections of the orbital angular momentum ${\bf l}$ and spin
$\mbox{\boldmath$\sigma$}$ across the plane of motion.

In the Hartree-Fock (HF) approximation the equation for the HF
wavefunctions $\psi$ reads:
\begin{equation}
   [H_B+V_i]\psi_i({\bf r}\sigma)+
   \sum_{\sigma'}\int\! d{\bf r}'\,
   \Delta_i({\bf r}\sigma,{\bf r}'\sigma')
   \psi_i({\bf r}'\sigma')=
   \varepsilon_i\psi_i({\bf r}\sigma),
\label{h-f}
\end{equation}
where $H_B$ is the hamiltonian of a single (non-interacting) electron
for the field $B$ defined in Eq.(\ref{ham}), $V_i$ denotes the Hartree
potential:
\begin{equation}
   V_i={e^2\over\epsilon}\int\! d{\bf r}'\,
   {n_i({\bf r}')\over|{\bf r}'-{\bf r}|},
\end{equation}
with:
\begin{equation}
   n_i({\bf r})=\sum_\sigma{\sum_j}'|\psi_j({\bf r}\sigma)|^2,
\end{equation}
and $\Delta_i$ is the Fock correction:
\begin{equation}
   \Delta_i({\bf r}\sigma,{\bf r}'\sigma')=
   -{e^2\over\epsilon}\delta_{\sigma'\sigma}{\sum_j}'
   {\psi^*_j({\bf r}'\sigma')\psi_j({\bf r}\sigma)
   \over|{\bf r}'-{\bf r}|}.
\end{equation}
Introducing the exchange operator $G$ as:
\begin{equation}
   G_i\psi_i({\bf r}\sigma)
   =\sum_{\sigma'}\int\! d{\bf r}'\,
   \Delta_i({\bf r}\sigma,{\bf r}'\sigma')
   \psi_i({\bf r}'\sigma'),
\label{opg}
\end{equation}
the HF equations (Eq.\ref{h-f}) can be written in a compact form:
\begin{equation}
   [H_B+V_i+G_i]\,\psi_i({\bf r}\sigma)
   =\varepsilon_i\psi_i({\bf r}\sigma).
\end{equation}

Following the work by Shikin {\sl et al.} \cite{shikin} we employ
here the approximate formula for the charge density in the parabolic
dot (of confining frequency $\omega_0$), obtained within the classical
regime and therefore applicable for large numbers of electrons $N$:
\begin{equation}
  n(r)=\left\{ \begin{array}{rl}
  n(0){1\over R}\sqrt{R^2-r^2}&\mbox{for $r\le R$}\\
  0&\mbox{for $r>R$}\end{array}\right.,
\label{chd}
\end{equation}
with the charge density in the center $n(0)=3N/2\pi R^2$.
The dot radius $R$, for the classical system given explicitely in
Ref. \onlinecite{shikin} will be calculated here in a variational
manner (from the minimum energy condition) in order to account for
the quantum corrections.
Using the formula (\ref{chd}) the Hartree potential is calculated:
\begin{equation}
   V_R(r)
   ={3\pi Ne^2\over4\epsilon R}\left(1-{r^2\over2R^2}\right)
\label{vr}
\end{equation}
(we take this form also for $r>R$).

\section{Quantum dot at zero magnetic field}

\subsection{Single-electron states}

Let us first consider the case of zero magnetic field.
Neglecting for the moment the exchange term (which will be included
later as a perturbation) we obtain the following Hartree equations:
\begin{equation}
   \left(-{\hbar^2\over2m^*}\Delta
   +{1\over2}m^*\Omega_0'^2r^2
   +\alpha l\sigma\right)\psi_i({\bf r}\sigma)
   =\left(\varepsilon_i-{3\pi Ne^2\over4\epsilon R}\right)
   \psi_i({\bf r}\sigma),
\label{har-zf}
\end{equation}
where we use the effective frequency, renormalized by the
Hartree term (Eq.\ref{vr}):
\begin{equation}
   \Omega_0'^2=\omega_0^2-{3\pi Ne^2\over4\epsilon m^*R^3}.
\label{ef}
\end{equation}
Eq.(\ref{har-zf}) can be solved analitically and the obtained
eigen-states are:
\begin{equation}
   \psi_i({\bf r}\sigma)=\psi_{nm\sigma}({\bf r}\sigma)
   =\phi_m(\theta)R_{nm}(r)\chi_{\sigma},
\label{efun1}
\end{equation}
with the spin eigen-function $\chi_{\sigma}$
(with eigen-values $\sigma=\pm\hbar/2$), the angular
wavefunction of the angular momentum eigen-value $m$:
\begin{equation}
   \phi_m(\theta)={1\over\sqrt{2\pi}}e^{im\theta}
\end{equation}
and the orbital wavefunction:
\begin{equation}
   R_{nm}(r)
   ={\sqrt{2}\over l_0'}\sqrt{n_r!\over(n_r+|m|)!}
   \left({r\over l_0'}\right)^{|m|}e^{-r^2/2l_0'^2}
   L^{|m|}_{n_r}\left({r^2\over l_0'^2}\right).
\end{equation}
In the above $L_{n_r}^{|m|}$ are the Laguerre polynomials:
\begin{equation}
   L_{n_r}^{|m|}(z)={1\over m!}z^{-|m|}
   e^z{d^{n_r}\over dz^{n_r}}(z^{n_r+|m|}e^{-z}),
\end{equation}
$l_0'=\sqrt{\hbar/m\Omega_0'}$ is the characteristic length,
$n=0,1,2,\ldots$ is the principal quantum number,
$m$ is the azimuthal quantum number ($|m|\le n$ and the parity
of $m$ is the same as that of $n$),
and $n_r={n-|m|\over2}$ is the radial quantum number.

The eigen-energies associated with the eigen-functions
$\psi_{nm\sigma}$ are:
\begin{equation}
   \varepsilon_{nm\sigma}
   =\hbar\Omega_0'(n+1)+\alpha m\sigma+{3\pi Ne^2\over4\epsilon R}.
\end{equation}
In the absence of the spin-orbit interaction ($\alpha=0$) they form
degenerate shells labeled by $n$.
Non-vanishing $\alpha$ splits these shells into the doubly degenerate
sub-levels.

Often used is a complementary (Fock-Darwin) representation:
\begin{equation}
   \psi_i({\bf r}\sigma)=\psi_{n_+n_-\sigma}({\bf r}\sigma)
   =\phi_{n_+n_-}({\bf r})\chi_{\sigma},
\label{efun2}
\end{equation}
with $n_\pm=0,1,2,\ldots$
The two sets of quantum numbers: $[n,m]$ and $[n_+,n_-]$ are
connected by the simple relations: $n=n_++n_-$ and $m=n_+-n_-$.
The orbital part of $\psi_i$ is defined as:
\begin{equation}
   \psi_{n_+n_-}({\bf r})
   ={1\over\sqrt{2\pi}l_0'}
   {(a^+)^{n_+}(b^+)^{n_-}\over\sqrt{n_+!n_-!}}e^{-r^2/2l_0'^2},
\end{equation}
where the raising operators $a^+$ and $b^+$ are:
\begin{eqnarray}
   a^+&=&{1\over2i}\left[{x+iy\over l_0'}-l_0'\left(
   {\partial\over\partial x}+i{\partial\over\partial y}\right)\right]
\nonumber\\
   b^+&=&{1\over2}\left[{x-iy\over l_0'}-l_0'\left(
   {\partial\over\partial x}-i{\partial\over\partial y}\right)\right].
\end{eqnarray}
The eigen-energies labeled by $n_\pm$ and $\sigma$ are:
\begin{equation}
   \varepsilon_{n_+n_-\sigma}
   =\varepsilon_+(n_++{1\over2})+\varepsilon_-(n_-+{1\over2})
   +{3\pi Ne^2\over4\epsilon R},
\label{epms}
\end{equation}
where $\varepsilon_\pm=\hbar\Omega_0'\pm\alpha\sigma$.

\subsection{Many-electron ground state}

The ground state energy of the system in terms of the $N$ lowest
Hartree eigen-energies $\varepsilon_i$, with $i$ standing for the
composite index $[n,m,\sigma]$, reads:
\begin{equation}
   {\cal E}_0=\sum_{i=1}^N\varepsilon_i
   -{e^2\over2\epsilon}\int\! d{\bf r}\int\! d{\bf r}'
   \,{n(r)n(r')\over|{\bf r}-{\bf r}'|},
\label{mehgs}
\end{equation}
where the subtracted integral represents the direct Coulomb energy
of the system, counted twice in the summation of the Hartree energies
$\varepsilon_i$.
Introducing the Fermi energy $\varepsilon_F$ separating the occupied
and unoccupied Hartree energy levels in the ground state, and
calculating the self-interaction integral we arrive at the formula:
\begin{equation}
   {\cal E}_0
   =\sum_i\Theta(\varepsilon_F-\varepsilon_i)\,\varepsilon_i
   -{3\pi N^2e^2\over10\epsilon R}.
\label{gse}
\end{equation}
In the above $\Theta$ is the Heaviside function.
The Fermi energy is determined by imposing the fixed number of
electrons $N$:
\begin{equation}
   N=\sum_i\Theta(\varepsilon_F-\varepsilon_i).
\label{gsn}
\end{equation}

The details of calculating the Hartree energy ${\cal E}_0$ are given
in Appendix \ref{appA}, and here we shall only present the final result:
\begin{equation}
   {\cal E}_0
   ={9\pi N^2e^2\over20\epsilon R}
   +{2\over3}N^{3/2}\hbar\Omega_0'\sqrt{1-{\beta^2N\over36}}.
\label{heazmf}
\end{equation}

The radius of the dot $R$ can now be determined from the minimum
condition: $\partial{\cal E}_0/\partial R=0$, equivalent to the
equation:
\begin{equation}
   \omega_0^2={3\pi Ne^2\over4\epsilon R^3m^*}
   \left[1-{100a_B\over27\pi R}\left(1-{\beta^2N\over36}\right)\right],
\end{equation}
where $a_B=\epsilon\hbar^2/m^*e^2$ is the effective Bohr radius.
Confining our considerations to the case of the large number
of electrons, we can solve the above equation perturbatively
with respect to the small parameter $a_B/R\ll1$.
The zeroth order approximation $R_0$ can be written in the form:
\begin{equation}
   R_0^3={3\pi Ne^2\over4\epsilon m^*\omega_0^2}
\label{rzero}
\end{equation}
and coincides with the classical result obtained in Ref.
\onlinecite{shikin}.
Assuming the first order approximation in the form: $R=R_0(1+\delta)$,
the correction $\delta$ reads:
\begin{equation}
   \delta={100a_B\over81\pi R_0}\left(1-{\beta^2N\over36}\right).
\end{equation}
Disregarding higher corrections (non-linear in $\delta$), the
effective confining frequency $\Omega_0'$ can be found based
on its definition (\ref{ef}):
\begin{equation}
   \Omega_0'^2=\Omega_0^2\left(1-{\beta^2N\over36}\right),
\end{equation}
where we use the notation:
\begin{equation}
   \Omega_0^2={100a_B\over27\pi R_0}\omega_0^2.
\label{Om0}
\end{equation}
Finally, we arrive at the formula for the ground state energy
in the Hartree approximation:
\begin{equation}
   {\cal E}_0
   ={9\pi N^2e^2\over20\epsilon R_0}
   +{1\over3}\hbar\Omega_0\left(1-{\beta^2N\over36}\right)N^{3/2}.
\end{equation}
The first term in the above equation, calculated in Ref.
\onlinecite{shikin}, is the classical energy of $N$ interacting
electrons confined in a parabolic well:
\begin{equation}
   {9\pi N^2e^2\over20\epsilon R_0}
   =\int\! d{\bf r}\, n(r)\cdot{1\over2}m^*\omega_0^2r^2
   +{e^2\over2\epsilon}\int\! d{\bf r}\! \int d{\bf r}'
   \,{n(r)n(r')\over|{\bf r}-{\bf r}'|}.
\end{equation}
The second term is the quantum correction and splits into the
energy of oscillation with the frequency $\Omega_0$:
\begin{equation}
   {1\over3}\hbar\Omega_0N^{3/2}
   =\sum_i\Theta(\varepsilon_F-\varepsilon_i)
   \,{1\over2}m^*\Omega_0^2<i|r^2|i>
\end{equation}
and the spin-orbit interaction term.

Let us now include perturbatively the exchange interaction, neglected
so far in the Hartree approximation.
As the first order correction $\Delta{\cal E}$ we shall calculate the
average value of the exchange operator $G$, defined by Eq.(\ref{opg}),
in the Hartree ground state obtained without including the spin-orbit
interaction:
\begin{equation}
   \Delta{\cal E}
   =\sum_i\Theta(\varepsilon_F-\varepsilon_i)\,
   \left.<i|G_i|i>\right|_{\beta=0}.
\label{fewsoi}
\end{equation}
We have verified that the effect due to the spin-orbit interaction
is indeed negligible here:
for $\beta=0.3$ the correction to $\Delta\cal E$ does not exceed
0.15 meV per electron for $N<100$ (compare with the energy scale in
Fig.\ref{fig1}).
Similarly, the second order correction proved to be smaller than the
first order correction by the orders of magnitude (reaching merely
$10^{-2}$ meV), which is then a good approximation to the actual
exchange energy.

As shown in Appendix \ref{appC}, Eq.(\ref{fewsoi}) can be
conveniently written as:
\begin{equation}
   \Delta{\cal E}
   =-{4\sqrt{5}\over9\sqrt{3}}
   \left(1-{3\over4\sqrt{N}}\right)
   {N^{7/4}e^2\over\epsilon R_0}(1-\delta_0),
\label{ffe}
\end{equation}
where $\delta_0=\delta(\beta=0)$.

Thus we have obtained the total ground state energy $\cal E$ of the
system of $N$ electrons confined in a parabolic well, including the
kinetic energy, the direct and exchange Coulomb interaction, and the
spin-orbit coupling:
\begin{equation}
   {\cal E}={\cal E}_0+\Delta{\cal E}.
\end{equation}
In Fig.\ref{fig1} we present the average ground state energy per
electron $\varepsilon={\cal E}/N$ plotted as a function of the
number of electrons $N$.
The two curves corresponding to the parameter $\beta$ equal
0.3 and 0.6 are shown to be a reasonable interpolation between
the classical result by Shikin {\sl et al.} \cite{shikin} and
the experimental data by Ashoori {\sl et al.} \cite{ashoori}.

\subsection{Far infra-red absorption}

Let us now consider the selection rules for the optical
transitions of the system under the far infra-red (FIR) radiation.
Absorption of the FIR light, leading to the excitation of the
electron droplet, has been a powerful tool in the experimental
studies of quantum dots \cite{heitmann,alsmeier,demel,meurer}.

Since the wavelength of the FIR light is much larger that the
radius of the dot, one can use the dipole approximation for
describing the interaction between the light and electrons.
The probability of the optical transition between the initial
($i$) and final ($f$) states is proportional to the squared
matrix element of this interaction:
\begin{equation}
   d_{fi}^2\sim|<f|
   e{\bf E}\cdot\sum_i\Theta(\varepsilon_F-\varepsilon_i)
   \,{\bf r}_i|i>|^2,
\end{equation}
where ${\bf E}$ is the electric field, uniform over the volume
of the dot.
The dipole matrix element $d_{fi}$ vanishes unless there is a pair
of the HF states, one in the initial and the other in the final
many-electron state, with equal spins and different by unity in
each of the orbital quantum numbers [$n,m$]:
\begin{equation}
   \sigma^f=\sigma^i, \rule{1em}{0ex}
   |n^f-n^i|=1, \rule{1em}{0ex}
   |m^f-m^i|=1,
\label{srul}
\end{equation}
with all other corresponding HF states equal in the initial
and final state.
In other words, the absorption of a FIR photon leads to the excitation
of a single electron from its (initial) HF state to another (final) HF
state with the same spin $\sigma$ and the orbital quantum numbers changed
according to Eq.(\ref{srul}).
Translating these selection rules to the Fock-Darwin representation
we have:
\begin{eqnarray}
  &\phantom{\rm or:} \rule{1em}{0ex}
   \sigma^f=\sigma^i, \rule{1em}{0ex}
   n_+^f=n_+^i\pm1, \rule{1em}{0ex}
   n_-^f=n_-^i,&
   \nonumber\\
  &{\rm or:} \rule{1em}{0ex}
   \sigma^f=\sigma^i, \rule{1em}{0ex}
   n_-^f=n_-^i\pm1, \rule{1em}{0ex}
   n_+^f=n_+^i,&
\end{eqnarray}
i.e. the excited electron changes one of its orbital quantum numbers
[$n_+,n_-$] by unity.

The above selection rules lead to the splitting of the resonance
energy:
\begin{equation}
   {\cal E}^f-{\cal E}^i
   =\varepsilon_\pm=\hbar\Omega_0'\pm{1\over2}\alpha,
\end{equation}
The magnitude of this splitting $\alpha$ depends on the number of
electrons according to Eq.(\ref{esta}).

\section{Quantum dot in a magnetic field}

\subsection{Single-electron states}

Sketched in the previous section for the case of zero magnetic
field the procedure of minimization of the Hartree energy with
respect to the dot radius $R$, with later perturbative inclusion
of the exchange interaction, has been also carried out for nonzero
fields.
The explicit form of the Hartree equations including the presence
of a perpendicular magnetic field is:
\begin{equation}
   \left(-{\hbar^2\over2m^*}\Delta
   +{1\over2}m^*\left(\Omega_B'^2+{1\over4}\omega_c^2\right)r^2
   -{1\over2}\hbar\omega_cl
   -g\mu_B\sigma B+\alpha l\sigma
   \right)\psi_i({\bf r}\sigma)
   =\left(\varepsilon_i-{3\pi Ne^2\over4\epsilon R(B)}\right)
   \psi_i({\bf r}\sigma),
\label{har-nzf}
\end{equation}
where $\omega_c=eB/m^*c$ is the effective cyclotron frequency
and the zero field radius $R$ appearing in the definition of
$\Omega_B'$ (\ref{ef}) is now replaced by $R(B)$.
We shall also denote the total confining frequency by:
$\Omega'^2=\Omega_B'^2+\omega_c^2/4$, and its corresponding
characteristic length by: $l=\sqrt{\hbar/m^*\Omega'}$.

The eigen-functions of Eq.(\ref{har-nzf}) are of the
same form as given by Eqs.(\ref{efun1}) and (\ref{efun2}),
only with the characteristic length replaced by $l$.
The corresponding eigen-energies read:
\begin{equation}
   \varepsilon_{nm\sigma}
   =\hbar\Omega'(n+1)
   -{1\over2}\hbar\omega_cm
   -g\mu_B\sigma B
   +\alpha m\sigma
   +{3\pi Ne^2\over4\epsilon R(B)},
\label{elev1}
\end{equation}
or, in the other representation:
\begin{equation}
   \varepsilon_{n_+n_-\sigma}
   =\varepsilon_+(n_++{1\over2})+\varepsilon_-(n_-+{1\over2})
   -g\mu_B\sigma B
   +{3\pi Ne^2\over4\epsilon R},
\label{elev2}
\end{equation}
with $\varepsilon_\pm=\Omega_0'\pm(\hbar\omega_c/2+\alpha\sigma)$.
Since the Zeeman splitting is rather small for GaAs ($g\sim{1\over2}$
yielding $g\mu_B\sim0.05$ meV/T), which is the most common material used
for the quantum dots, we shall neglect it in the further considerations.

Including the magnetic field leads to the possibility of crossings
between different energy levels $\varepsilon_i$.
Whether the two close levels $\varepsilon_1$ and $\varepsilon_2$
actually cross, or their crossing is forbidden, depends on vanishing
of the off-diagonal matrix element of the operator describing the
change of the hamiltonian due to small change of the field.
Thus the condition for the allowed level-crossing is:
\begin{equation}
   <1|{\partial H\over\partial B}|2>\equiv0.
\label{cond}
\end{equation}
The operator $\partial H/\partial B$ commutes with the spin and
inversion (${\bf r}\rightarrow-{\bf r}$) operators.
Its commutation with the angular momentum requires assumed here
circular symmetry of the confining potential.
Hence, while for the states of non-equal quantum numbers $n$ and
$\sigma$ we have the condition (\ref{cond}) guaranteed, for a pair
of states different only in $m$ in general it is no longer true.
This leads to the anti-crossing of levels, which can be taken into
account by changing the formulae for eigen-energies:
\begin{equation}
   \varepsilon_{nm\sigma}
   =\hbar\Omega'(n+1)
   +m\left|{1\over2}\hbar\omega_c+\alpha\sigma\right|
   +{3\pi Ne^2\over4\epsilon R(B)}.
\end{equation}
Analogously we have to modify the definition of a pair of energies
$\varepsilon_+$ and $\varepsilon_-$, appearing in Eq.(\ref{elev2}),
into: $\varepsilon_\pm=\Omega_0'\pm|\hbar\omega_c/2+\alpha\sigma|$.
Let us underline that this rearrangement of levels is a perturbative
approximation, beyond the Hartree-Fock approach.

\subsection{Many-electron ground state}

The many-electron Hartree ground state in the magnetic field
${\cal E}_0(B)$ is defined analogously as in Eq.(\ref{mehgs}).
Using the procedure sketched in Appendix \ref{appD} one can
bring it to the form:
\begin{equation}
   {\cal E}_0(B)
   ={9\pi N^2e^2\over20\epsilon R_0}
   +{2\over3}N^{3/2}\hbar
   \left(\omega_0^2u_B-{3\pi Ne^2\over4\epsilon m^*R(B)^3}\right)^{1/2}
   \sqrt{1-{\beta^2f_B^2N\over36}},
\label{heiamf}
\end{equation}
with the functions $f_B$ and $u_B$ defined in Appendix \ref{appD}
by Eq.(\ref{deffB}) and Eq.(\ref{defuB}), respectively.

Analogously as for the zero magnetic field, the ground state radius
of the dot $R(B)$ can be found from the minimum energy condition:
\begin{equation}
   {\partial{\cal E}_0(B)\over\partial R(B)}=0
\end{equation}
which resolves into the equation:
\begin{equation}
   \omega_0^2u_B={3\pi Ne^2\over4\epsilon m^*R(B)^3}
   \left[1+{100a_B\over27\pi R(B)}
   \left(1-{\beta^2f_B^2N\over36}\right)\right].
\end{equation}
In the zeroth order approximation we obtain:
\begin{equation}
   R_0(B)^3={3\pi Ne^2\over4\epsilon m^*\omega_0^2u_B}
   ={1\over u_B}R_0^3,
\end{equation}
where $R_0$ is given by Eq.(\ref{rzero}).
The first order correction $\delta(B)$ defined as:
$R(B)=R_0(B)(1+\delta(B))$ reads:
\begin{equation}
   \delta(B)={100a_B\over81\pi R_0(B)}
   \left(1-{\beta^2f_B^2N\over36}\right).
\end{equation}
In Fig.\ref{fig2} we have drawn the radius as a function of the field.
The dependence is fairly weak.
When the magnetic field is increased, at low fields the radius also
increases, and later, at higher fields, slightly falls down.
The initial increase of the radius in the rising field is an effect
due to the spin-orbit interaction, and vanishes for $\beta=0$.

Finally, the ground state energy in the Hartree approximation
can be found in the form:
\begin{equation}
   {\cal E}_0(B)
   ={9\pi N^2e^2\over20\epsilon R_0(B)}
   +{1\over3}u_B^{2/3}\hbar\Omega_0
   \left(1-{\beta^2f_B^2N\over36}\right)N^{3/2}.
\end{equation}

Calculating the correction due to the exchange energy in the
obtained above Hartree ground state is far more complicated
for non-zero magnetic fields.
Therefore the hypothesis is used, according to which the kind
of dependence of the exchange energy on the number of particles
is not affected by the presence of the field \cite{vonsovskij}.
As a result we obtain the following formula for the exchange
energy:
\begin{equation}
   \Delta{\cal E}(B)
   =-{4\sqrt{5}\over9\sqrt{3}}
   \left(1-{3\over4\sqrt{N}}\right)
   {N^{7/4}e^2\over\epsilon R_0(B)}(1-\delta_0(B)),
\end{equation}
where $\delta_0(B)=\delta(B;\beta=0)$.
Thus the total ground state energy within our approach reads:
\begin{equation}
   {\cal E}(B)={\cal E}_0(B)+\Delta{\cal E}(B).
\end{equation}
and the average energy per particle is $\varepsilon(B)={\cal E}(B)/N$.

In Fig.\ref{fig3} we drew the magnetic field evolution of the average
ground state energy per electron $\varepsilon(B)={\cal E}(B)/N$.
The three frames correspond to the parameter $\beta$ equal 0.0
(no spin-orbit coupling), 0.3 and 0.6.
We find the qualitative agreement between our curves and the data
reported by Ashoori {\sl et al.}, obtained in the single electron
capacitance spectroscopy (SECS) experiment \cite{ashoori}.
Comparing the curves in the three frames one can conclude that
including the spin-orbit interaction brings the model curves fairly
close to the measured behavior (the fitting is particularly good
for $\beta=0.3$).

\subsection{Far infra-red absorption}

Let us now discuss the FIR absorption under the magnetic field.
Since the magnetic field does not affect the structure of the HF
wavefunctions, the selection rules (\ref{srul}) remain unchanged
and the transition energies are:
\begin{equation}
   {\cal E}^f(B)-{\cal E}^i(B)
   =\varepsilon_\pm=\hbar\Omega'\pm
   \left|{1\over2}\hbar\omega_c\pm{1\over2}\alpha(B)\right|
\end{equation}
and we deal with four resonance branches.

In Fig.\ref{fig4} we compare the dependence of the FIR resonance
energies obtained within our model with that reported by Demel
{\sl et al.} \cite{demel}.
Assuming that the experimentally observed higher mode is due to
the spin-orbit interaction as presented here, we again managed to
find a good agreement for $\beta=0.3$.
Particularly, the zero-field lower resonance energy ($\sim2.8$ meV),
the magnetic field at which the anti-crossing occurs ($\sim1$ Tesla),
and the energy at this crossing ($\sim3.9$ meV), seem to be fit very well.
A gap separating the anti-crossing levels, observed in the experiment,
is probably due to a slight anisotropy of the confining potential.

\section{Conclusions}

The self-consistent theory of a many-electron quantum dot has been
developed with the inclusion of the electron-electron (Coulomb)
and spin-orbit interactions.
Included here quantum corrections to the ground state energy of the
system improve the classical result by Shikin {\sl et al.} \cite{shikin}
and compare well with the experiments by Demel {\sl et al.} \cite{demel}
and Ashoori {\sl et al} \cite{ashoori}.
The spin-orbit interaction is shown to have a strong effect on the
electronic structure of the system, leading to the splitting of the
resonance energy in the far infra-red (FIR) absorption.
Predicted here anti-crossing of the FIR absorption modes in the
magnetic field describes very well the similar behavior reported
by Demel {\sl et al} \cite{demel}.

The critical magnetic field, at which the FIR modes cross and a bump
in the ground state energy occurs, is around 1 T, which agrees with
the experiments for GaAs \cite{demel,ashoori}, and on the other hand
with the numerical calculations for a two-electron InSb dot by
Darnhofer and R\"ossler \cite{darnhofer}.

\section*{Acknowledgment}

The work was supported by the KBN grants PB 674/P03/96/10 and
PB 1152/P03/94/06.
We also wish to thank Dr. Pawel Hawrylak (IMS NRC Ottawa) for
useful discussions and comments.

\appendix

\section{Hartree energy at zero magnetic field}
\label{appA}

In order to perform the summations over the occupied states in
Eqs.(\ref{gse}--\ref{gsn}), it is convenient to introduce the non-zero
temperature of the electrons $T$, and eventually find the limit for
$T\rightarrow0$.
The temperature leads to a replacement of the sharp Heaviside function
in Eqs.(\ref{gse}--\ref{gsn}) by the smooth Fermi distribution function:
\begin{equation}
   n_s(\varepsilon_+n_++\varepsilon_-n_-)=\left(
   {1+\exp{\varepsilon_+n_++\varepsilon_-n_--\mu_s\over k_BT}}
   \right)^{-1},
\end{equation}
where $\mu_s=\mu-{1\over2}(\varepsilon_++\varepsilon_-)$ and
$\mu$ is the chemical potential.

In order to conveniently hide for the moment the constant terms in
Eqs.(\ref{epms}) and (\ref{gse}) we introduce the primed energy symbol:
\begin{equation}
   {\cal E}'={\cal E}
   -N\cdot{3\pi Ne^2\over4\epsilon R}
   +{3\pi Ne^2\over10\epsilon R}
   ={\cal E}+{9\pi Ne^2\over20\epsilon R}
\label{prime}
\end{equation}
and rewrite Eqs.(\ref{gse}--\ref{gsn}) in the form:
\begin{eqnarray}
   N&=&\sum_{n_+n_-\sigma}
   n_s(\varepsilon_+n_++\varepsilon_-n_-),
\label{gsnt} \\
   {\cal E}'(T)&=&
   \sum_{n_+n_-\sigma}
   (\varepsilon_+n_++\varepsilon_-n_-)\cdot
   n_s(\varepsilon_+n_++\varepsilon_-n_-)+
   N{\varepsilon_++\varepsilon_-\over2}.
\label{gset}
\end{eqnarray}
In order to find the thermodynamically stable state we shall further
introduce the following thermodynamical potential:
\begin{equation}
   \Phi=\sum_{n_+n_-\sigma}\phi_s(\varepsilon_+n_++\varepsilon_-n_-),
\end{equation}
where we define:
$\phi_s(\varepsilon)=-k_BT\ln(1+\exp{\mu_s-\varepsilon\over k_BT})$.
At low temperatures the minimization procedure with respect to $\Phi$
is equivalent to finding the ground state, as we have:
\begin{equation}
   {\cal E}'(T)=\Phi+\mu N-TS
\end{equation}
(with the entropy $S=\partial\Phi/\partial T$) and:
\begin{equation}
   {\cal E}'_0=\Phi_0+\varepsilon_F N,
\label{limit}
\end{equation}
where: $\Phi_0=\lim_{T\rightarrow 0}\Phi$ and
$\varepsilon_F=\lim_{T\rightarrow 0}\mu$.

Let us now introduce the Laplace transformations:
\begin{eqnarray}
   n_s(\varepsilon)&=&{1\over2\pi i}
   \int_{c-i\infty}^{c+i\infty}dp\,\tilde{n}_s(p)e^{p\varepsilon},\;\;
   \tilde{n}_s(p)=
   \int_0^\infty d\varepsilon\,n_s(\varepsilon)e^{-\varepsilon p},
   \nonumber\\
   \phi_s(\varepsilon)&=&{1\over2\pi i}
   \int_{c-i\infty}^{c+i\infty}dp\,\tilde{\phi}_s(p)e^{p\varepsilon},\;\;
   \tilde{\phi}_s(p)=
   \int_0^\infty d\varepsilon\,\phi_s(\varepsilon)e^{-\varepsilon p}.
\end{eqnarray}
Using these expressions we can rewrite Eqs.(\ref{gsnt}--\ref{gset})
as follows:
\begin{eqnarray}
   N&=&\sum_\sigma \int_0^\infty d\varepsilon z_0(\varepsilon)
   \left(-{\partial n_s\over\partial \varepsilon}\right),
   \\
   \Phi&=&\sum_\sigma \int_0^\infty d\varepsilon z_1(\varepsilon)
   \left(-\frac{\partial n_s}{\partial \varepsilon}\right),
\end{eqnarray}
where
\begin{eqnarray}
   z_0(\varepsilon)&=&
   {1\over2\pi i}\int_{c-i\infty}^{c+i\infty} dp\,
   {e^{\varepsilon p}\over
   p(1-e^{-p\varepsilon_+})(1-e^{-p\varepsilon_-})},
   \\
   z_1(\varepsilon)&=&
   {1\over2\pi i}\int_{c-i\infty}^{c+i\infty} dp\,
   {e^{\varepsilon p}\over
   p^2(1-e^{-p\varepsilon_+})(1-e^{-p\varepsilon_-})}.
\end{eqnarray}
We chose the constant $c$ in such a way that all the singular points
of the subintegral lie on the left-hand side of the integral contour
(the contour encloses all singularities inside).
Since  $-\partial n_s/\partial \varepsilon$ tends to the Dirac's
delta for $T\rightarrow0$, one finds that:
\begin{eqnarray}
   N&=&
   {1\over2}\sum_\sigma \left\{
   \left[ \frac{\mu_{0s}^2}{\varepsilon_+\varepsilon_-}+
          \mu_{0s}\left({1\over\varepsilon_+}+{1\over\varepsilon_-}\right)+
          {1\over2}\right]\right.
\label{thrthr}
\nonumber\\
   &+&
   \left[P_1\left({\mu_{0s}\over\varepsilon_+}\right)+
         P_1\left({\mu_{0s}\over\varepsilon_-}\right)\right]-
   {\varepsilon_+\over\varepsilon_-}
   \left[P_2\left({\mu_{0s}\over\varepsilon_+}\right)-{1\over12}\right]
\nonumber\\
   &-& \left.
   2{\varepsilon_-\over\varepsilon_+}
   \left[P_2\left({\mu_{0s}\over\varepsilon_-}\right)-{1\over12}\right]+
   2F_0\left({\mu_{0s}\over\varepsilon_+},
   {\mu_{0s}\over\varepsilon_-}\right)\right\},
\label{nfiz}
\end{eqnarray}
where $\mu_{0s}=\varepsilon_F-{1\over2}(\varepsilon_++\varepsilon_-)$,
and:
\begin{eqnarray}
   \Phi_0&=&
   -{1\over6}\sum_\sigma\left\{
   \varepsilon_F\left[{\mu_{0s}^2\over\varepsilon_+\varepsilon_-}
   +\mu_{0s}\left({1\over\varepsilon_+}+{1\over\varepsilon_-}\right)
   +{1\over2}\right]\right.
\nonumber\\
   &-&
   3\left[\varepsilon_+P_2\left({\mu_{0s}\over\varepsilon_+}\right)
   +\varepsilon_-P_2\left({\mu_{0s}\over\varepsilon_-}\right)\right]
   -6\left[{\varepsilon_+^2\over\varepsilon_-}
   P_3\left({\mu_{0s}\over\varepsilon_+}\right)+
   {\varepsilon_-^2\over\varepsilon_+}
   P_3\left({\mu_{0s}\over\varepsilon_-}\right)\right]
\nonumber\\
   &-& \left.
   6\sqrt{\varepsilon_+\varepsilon_-}
   F_1\left({\mu_{0s}\over\varepsilon_+},
            {\mu_{0s}\over\varepsilon_-}\right)\right\}.
\label{fiz}
\end{eqnarray}
In the above formulae we have introduced the following notation:
\begin{eqnarray}
   P_{2j}(x)&=&\sum_{i=1}^\infty
   {\cos(2\pi ix)\over2^{2j-1}i^{2j}\pi^{2j}},
   \nonumber \\
   P_{2j+1}(x)&=&\sum_{i=1}^\infty
   {\sin (2\pi ix)\over2^{2j}i^{2j+1}\pi^{2j+1}},
   \nonumber \\
   F_0(x,y)&=&-{\varepsilon_+\varepsilon_-\over\pi^2}
   {\sum_{i,j=1}^\infty}'
   {\cos(2\pi ix)-\cos(2\pi jy)\over
   (i\varepsilon_-)^2-(j\varepsilon_+)^2}
   \nonumber \\
   F_1(x,y)&=&{\sqrt{\varepsilon_+\varepsilon_-}\over\pi^2}
   {\sum_{i,j=1}^\infty}'
   {{\varepsilon_+\over2\pi i}\sin(2\pi ix)-
    {\varepsilon_-\over2\pi j}\sin(2\pi jy)
    \over(i\varepsilon_-)^2-(j\varepsilon_+)^2}.
\end{eqnarray}
The above functions are periodic with the period 1.
In the relevant range of domain the absolute value of each function
is less then unity.
Moreover, note that for $0<x<1$
\begin{eqnarray}
   P_1(x)\simeq -x+{1\over2} \;\; &\rightarrow& \;\;
   |P_1(x)|\leq {1\over2},
   \nonumber \\
   P_2(x)\simeq {x^2\over2}-{x\over2}+{1\over12} \;\; &\rightarrow& \;\;
   |P_2(x)|\leq {1\over12},
   \nonumber \\
   P_3(x)\simeq {x^3\over6}-{x^2\over4}+{x\over12} \;\; &\rightarrow& \;\;
   |P_3(x)|\leq 0.009.
\end{eqnarray}
The energy of the system ${\cal{E}}_0'$ one can find from Eq.(\ref{fiz})
using $\varepsilon_F(N)$ determined from Eq.(\ref{nfiz}).
Taking into account that $\varepsilon_\pm=\Omega_0'\pm\alpha\sigma$
and introducing:
$\nu_F=\varepsilon_F/\sqrt{\varepsilon_+\varepsilon_-}$,
Eq.(\ref{nfiz}) can be rewritten as follows:
\begin{eqnarray}
   &N+{1\over2}\,
   {(\hbar\Omega_0')^2+(\alpha_0/2)^2\over
    (\hbar\Omega_0')^2-(\alpha_0/2)^2}
   =\nu_F^2+{1\over2}\sum_\sigma
   \left\{
   \left[P_1\left({\mu_{0s}\over\varepsilon_+}\right)+
         P_1\left({\mu_{0s}\over\varepsilon_-}\right)\right]
   \right.&
   \nonumber \\
   &\left.
   -2{\varepsilon_+\over\varepsilon_-}
   \left[P_2\left({\mu_{0s}\over\varepsilon_+}\right)-{1\over12}\right]
   -2{\varepsilon_-\over\varepsilon_+}
   \left[P_2\left({\mu_{0s}\over\varepsilon_-}\right)-{1\over12}\right]
   +2F_0\left({\mu_{0s}\over\varepsilon_+},
              {\mu_{0s}\over\varepsilon_-}\right)\right\}.&
\label{follows}
\end{eqnarray}
The absolute values of all periodic functions are ranged here by unity.
Hence, for $N\gg1$, we note that all the terms under the summation on
the right-hand side of the above equation are small compared to $N$
if only:
\begin{equation}
   \left|\,2\,{\hbar\Omega_0'+\alpha/2\over\hbar\Omega_0'-\alpha/2}
   \left[P_2\left({\mu_{0s}\over\hbar\Omega_0'+\alpha/2}\right)-
   {1\over12}\right]\,\right|\leq1
   \rule{1em}{0ex}
   {\rm or:}
   \rule{1em}{0ex}
   \alpha\leq{6\over5}\hbar\Omega_0'.
\label{crit}
\end{equation}
Provided this condition one can use the perturbation method and look
for the solution in the form of the following series:
\begin{equation}
 \nu_F=\nu_{0F}+\nu_{1F}+\nu_{2F}+\ldots\;,
\end{equation}
where: $|\nu_{0F}|\gg|\nu_{1F}|\gg|\nu_{2F}|\gg\ldots$
From Eq.(\ref{follows}) we find:
\begin{equation}
   \nu_{0F}=\sqrt{N+{1\over2}\,
   {(\hbar\Omega_0')^2+(\alpha/2)^2\over
    (\hbar \Omega_0')^2-(\alpha/2)^2}}\sim\sqrt{N}
\end{equation}
and, using: $\mu_{0s}^o=\varepsilon_{0F}-\hbar\Omega_0'=
\nu_{0F}\sqrt{(\hbar\Omega_0')^2-(\alpha/2)^2}-\hbar\Omega_0'$:
\begin{eqnarray}
   \nu_{1F}=-{1\over4\nu_{0F}}\sum_\sigma
   \left\{
   P_1\left({\mu_{0s}^o\over\varepsilon_+}\right)+
   P_1\left({\mu_{0s}^o\over\varepsilon_-}\right)+
  2F_0\left({\mu_{0s}^o\over\varepsilon_+},
            {\mu_{0s}^o\over\varepsilon_+}\right)\right.
   \nonumber \\ \left.
   -2{\varepsilon_+\over\varepsilon_-}
   \left[P_2 \left({\mu_{0s}^o\over\varepsilon_+}\right)
         -{1\over12}\right]
   -2 {\varepsilon_-\over\varepsilon_+}
   \left[ P_2\left({\mu_{0s}^o\over\varepsilon_-}\right)
         -{1\over12}\right]\right\}\sim{1\over\sqrt{N}}.
\label{nuof}
\end{eqnarray}
Using the above expressions for $\nu_F$ we find that (for $N\gg1$)
\begin{eqnarray}
   \varepsilon_F&=&\nu_{0F}\sqrt{(\hbar\Omega_0')^2-(\alpha/2)^2}
   \left[1+{\nu_{1F}\over\nu_{0F}}  +O(N^{-3/2})\right],
   \\
   \Phi_0&=&-{N\over3}\nu_{0F}\sqrt{(\hbar\Omega_0')^2-(\alpha/2)^2}
   \left[1+3{\nu_{1F}\over\nu_{0F}}  +O(N^{-3/2})\right],
   \\
   {\cal{E}}_0'&=&\Phi_0+N\varepsilon_F=
   {2N\over3} \nu_{0F} \sqrt{(\hbar\Omega_0')^2-(\alpha/2)^2}
   [1+O(N^{-3/2})].
\label{finale}
\end{eqnarray}
The main non-oscillating term in the above formula for $\Phi_0$ is
of the order of $N\sqrt{N}$ while the oscillating terms are at most
of the order of $\sqrt{N}$.
In the formula for ${\cal{E}}_0'$ the non-oscillating term is of
the order of $N\sqrt{N}$ and the oscillating terms are at most of
the order of unity.

The estimation of the magnitude of spin-orbit coupling constant
$\alpha$ is given in Appendix \ref{appB}.
Using Eq.(\ref{esta}), the final form of the formula for the Hartree
energy, given explicitely in Eq.(\ref{heazmf}), can be now obtained
from Eq.(\ref{finale}) by substituting Eq.(\ref{esta}) and shifting
${\cal E}_0'$ back by the previously omitted constant (according to
Eq.\ref{prime}).

\section{Spin-orbit coupling constant at zero magnetic field}
\label{appB}

According to Eq.(\ref{soc}) one needs to estimate the average
self-consistent field $<\delta U>$ acting on an electron.
The energy of a classical particle in the two-dimensional potential
${1\over2}m^*\Omega_0'^2r^2$, moving on the orbit of radius $r$, is:
\begin{equation}
   \delta U={1\over2}m^*\Omega_0'^2r^2+{1\over2}\Omega_0'l.
\end{equation}
Substituting the classical variables by the respective operators
and averaging $\delta U$ first over the quantum state $[n,m,\sigma]$:
\begin{equation}
   <nm\sigma|\delta U|nm\sigma>={1\over2}\hbar\Omega_0'(n+m+1)
\end{equation}
and then over all occupied states, one can obtain the following formula:
\begin{equation}
   <\delta U>={1\over2N}\hbar\Omega_0'
   \sum_{nm\sigma}(n+m+1)\cdot n_s(n\hbar\Omega_0'),
\label{averso}
\end{equation}
where the spin-orbit energy under the distribution function $n_s$
has been neglected.
One can now use the following relation:
\begin{equation}
   \sum_{nm\sigma}(n+m+1)\cdot n_s(nx+my)=
   \left({\partial\Phi_0\over\partial x}\right)_{x=\varepsilon_F}+
   \left({\partial\Phi_0\over\partial y}\right)_{y=\varepsilon_F},
\label{useful}
\end{equation}
and finally arrive at the formula:
\begin{equation}
   \alpha={1\over3}\beta\hbar\Omega_0'^2\sqrt{N}.
\label{esta}
\end{equation}
Based on the above equation, the second criterion in Eq.(\ref{crit})
can be expressed as:
\begin{equation}
   \sqrt{N}<{18\over5\beta},
\end{equation}
which gives $N<144$ for $\beta=0.3$.

\section{Fock energy at zero magnetic field}
\label{appC}

Eq.(\ref{fewsoi}) can be written as:
\begin{equation}
   \Delta{\cal E}
   =-{2e^2\over\epsilon}\sum_{n,n'}{\sum_{m,m'}}'
   I_{nm}^{n'm'}n_{0s}(n\hbar\Omega_0)n_{0s}(n'\hbar\Omega_0),
\label{sixtwo}
\end{equation}
where the prime in the second sum excludes terms with $m=m'$, and:
\begin{eqnarray}
   I_{n,m}^{n'm'}&=&
   \int\! d{\bf r}\,\int\! d{\bf r}'\,
   {\psi^*_{nm}({\bf r})\psi_{n'm'}({\bf r})
   \psi^*_{n'm'}({\bf r}')\psi_{nm}({\bf r}')
   \over |{\bf r}-{\bf r}'|}
   \nonumber \\ &=&
   {1\over l_0'}\int_0^\infty\! dx\,x\,\int_0^\infty\! dx'\,x\,
   \int_0^{2\pi} {d\theta\over 2\pi}
   e^{-i(m-m')\theta}{R_{nm}(x)R_{n'm'}(x)R_{nm}(x')R_{n'm'}(x')
   \over \sqrt{x^2-2xx'\cos\theta +x{'}^2}}.
\end{eqnarray}
Expanding the subintegral into the Legendre polynomials and integrating
over $\theta$ one obtains:
\begin{eqnarray}
   I_{n,m}^{n'm'}&=&{1\over l_0'}
   \int_0^\infty \! dx\,x^2R_{nm}(x)R_{n'm'}(x)
   \nonumber \\
   &\times&\left\{
   \int_0^1\! dt\,tR_{nm}(tx)R_{n'm'}(tx)t^{|\Delta m|}
   \sum_{k=0}^\infty t^{2k}a_{k+|\Delta m|}a_k \right.
   \nonumber \\
   &+&\left.\;
   \int_1^\infty dtR_{nm}(tx)R_{n'm'}(tx)t^{-|\Delta m|}
   \sum_{k=0}^\infty t^{-2k}a_{k+|\Delta m|}a_k \right\},
\label{sixthr}
\end{eqnarray}
where $\Delta m =m -m'$ and $a_k={(2k-1)!!\over(2k)!!}$.
The integral $I_{n,m}^{n'm'}$ decays rapidly for
$\Delta m\rightarrow\infty$ and hence we can cut off the summation
over $m'$ in Eq.(\ref{sixtwo}) keeping only the terms with $m'=m\pm1$.
In the integral over $x$ we deal with a subintegral with factor
$e^{-x^2}$ and thus the most contributes the region around $x=1$.
Provided that the function in braces (in Eq.\ref{sixthr}) is smooth
with respect to $x$, it can be replaced by its value at $x=1$.
Moreover, since $R_{nm}(t)R_{n'm'}(t)\sim e^{-t^2}$
the second term in the braces can be neglected as small in comparison
with the first one.
Hence, for $|\Delta m|=1$ the integral $I_{n,m}^{n'm'}$ attains the form:
\begin{eqnarray}
   I_{n,m}^{n'm'}&\simeq&
   {1\over l_0'}\int_0^\infty\! dx\,x^2 R_{nm}(x)R_{n'm'}(x)
    \int_0^1\!dt\,t^2 R_{nm}(t)R_{n'm'}(t)
    \left({1\over2}+{3\over16}t^2+\ldots\right)
    \nonumber \\
    &\simeq&{1\over3l_0'}\left(\int_0^\infty\! dx\,x^2
    R_{nm}(x)R_{n'm'}(x)\right)^2,
\end{eqnarray}
where we have taken into account the relation:
\begin{equation}
   \int_0^1\! dt\,t^2R_{nm}(t)R_{n'm'}(t)
   \simeq {2\over3}\int_0^\infty\! dt\,t^2R_{nm}(t)R_{n'm'}(t).
\end{equation}
Since:
\begin{equation}
   \int_0^\infty\! dx\,x^2 R_{nm}(x)R_{n'm\pm1}(x)=
   \pm{m\over|m|}\left[\sqrt{1+{n\pm m\over2}}\,\delta_{n',n+1}
   -\sqrt{n\mp m\over2}\,\delta_{n',n-1}\right],
\end{equation}
we have:
\begin{equation}
   I_{nm,n'm\pm 1}={1\over3l_0'}\left[
   \left(1+{n\pm m\over2}\right)\delta_{n',n+1}+
   \left({n\mp m\over2}\right)\delta_{n',n-1}\right],
\end{equation}
and:
\begin{eqnarray}
   \Delta{\cal E}&=&-{2e^2\over\epsilon l_0'}
   \sum_{nm,n'm'}n_{0s}(n\hbar\Omega_0)n_{0s}(n'\hbar\Omega_0)\,
   (I_{nm,n'm-1}\delta_{m',m-1}+I_{nm,n'm+1}\delta_{m',m+1})
   \nonumber \\
   &=&-{4e^2\over3\epsilon l_0'}
   \left[\sum_{nm}(n+1)n_{0s}(n\hbar\Omega_0)-{N\over4}-{1\over2}\right].
\end{eqnarray}
Using Eq.(\ref{useful}) we finally find the correction due to
exchange interaction in the form of Eq.(\ref{ffe}).

\section{Hartree energy in a magnetic field}
\label{appD}

Similarly as in the case of the zero magnetic field (see Appendix
\ref{appA}) it is convenient to shift the Hartree energy
${\cal E}_0(B)$ and introduce:
\begin{equation}
   {\cal E}_0'(B)={\cal E}_0(B)-{9\pi N^2e^2\over20\epsilon R(B)},
\end{equation}
given by:
\begin{equation}
   {\cal E}_0'(B)=\sum_{n_+n_-\sigma}
   (\varepsilon_+n_++\varepsilon_-n_-)\cdot
   n_s(\varepsilon_+n_++\varepsilon_-n_-)+
   N{\varepsilon_++\varepsilon_-\over2}.
\end{equation}
In order to calculate ${\cal E}_0'(B)$ we further introduce the notation:
\begin{eqnarray}
   \nu_F^2&=&
   \varepsilon_F^2\cdot{(\hbar\Omega_B')^2-(\alpha(B)/2)^2\over
   [(\hbar\Omega_B')^2-(\alpha(B)/2)^2]^2-(\hbar\omega_c\alpha(B)/2)^2}
   \\
   \zeta^2&=&{[(\hbar\Omega_B')^2-(\alpha(B)/2)^2]\cdot
   [(\hbar\Omega_B')^2-(\hbar\omega_c/2)^2]\over
   [(\hbar\Omega_B')^2-(\alpha(B)/2)^2]^2-(\hbar\omega_c\alpha(B)/2)^2}.
\end{eqnarray}
and rewrite Eq.(\ref{thrthr}) as:
\begin{eqnarray}
   N-{1\over2}+\zeta^2&=&\nu_F^2+{1\over2}\sum_\sigma
   \left\{ \left[
   P_1\left({\mu_{0s}\over\varepsilon_+}\right)
  +P_1\left({\mu_{0s}\over\varepsilon_-}\right)\right]
  +2F_0\left({\mu_{0s}\over\varepsilon_+},
  {\mu_{0s}\over\varepsilon_-}\right)
   \right.
   \nonumber \\
   &-& \left.
   2{\varepsilon_+\over\varepsilon_-}
   \left[P_2\left({\mu_{0s}\over\varepsilon_+}\right)-{1\over12}\right]
  -2{\varepsilon_-\over\varepsilon_+}
   \left[P_2\left({\mu_{0s}\over\varepsilon_-}\right)-{1\over12}\right]
   \right\}.
\label{eigeig}
\end{eqnarray}
All the oscillating functions here are small compared to $N$ if:
\begin{equation}
   \left|2{\varepsilon_+\over\varepsilon_-}
   \left[P_2\left({\mu_{0s}\over\varepsilon_+}\right)-{1\over12}\right]
   \right|\ll N
\end{equation}
or:
\begin{eqnarray}
   &&\sqrt{(\hbar\Omega_B')^2+({1\over2}\hbar\omega_c)^2}
  +{1\over2}\hbar\omega_c
   \left|1+2{\alpha(B)\sigma\over\hbar\omega_c}\right|
   \nonumber \\
   &&\ll 4N\left[
   \sqrt{(\hbar\Omega_B')^2+({1\over2}\hbar\omega_c)^2}
  -{1\over2}\hbar\omega_c
   \left|1+2{\alpha(B)\sigma\over\hbar\omega_c}\right|\right].
\end{eqnarray}
At low magnetic fields, i.e. for ${1\over2}\hbar\omega_c\ll\alpha(B)$,
the corrections to $\Omega'$ and $\alpha$ due to the field are negligible
and the above condition is satisfied for sufficiently high $N$.
At strong magnetic fields, when ${1\over2}\hbar\omega_c\gg\alpha(B)$,
it takes the form:
\begin{equation}
   \sqrt{1+\left({\omega_c\over2\Omega_B'}\right)^2}
   +{\omega_c\over2\Omega_B'}\ll
   4N\left[\sqrt{1+\left({\omega_c\over2\Omega_B'}\right)^2}
   -{\omega_c\over2\Omega_B'}\right],
\end{equation}
which leads to:
\begin{equation}
   \left({\omega_c\over 2\Omega_B'}\right)^2\ll N.
\end{equation}
For example, taking the parameters for GaAs and $N=15$ the above
yields: $B\ll17$ T.
The solution of Eq.(\ref{eigeig}) can be represented as the series:
$\nu_F=\nu_{0F}+\nu_{1F}+\ldots$, where:
\begin{equation}
   \nu_{0F}=\sqrt{N-{1\over2}+\zeta^2}.
\end{equation}
The first order correction we find as:
\begin{eqnarray}
   \nu_{1F}=-{1\over4\nu_{0F}}\sum_\sigma && \left\{
   \left[P_1\left({\mu_{0s}^0\over\varepsilon_+}\right)
        +P_1\left({\mu_{0s}^0\over\varepsilon_-}\right)\right]
   +2F_0\left({\mu_{0s}^0\over\varepsilon_+},
   {\mu_{0s}^0\over\varepsilon_-}\right)
   \right.
   \nonumber \\ && \left. \;
   -2{\varepsilon_+\over\varepsilon_-}
   \left[P_2\left({\mu_{0s}^0\over\varepsilon_+}\right)-{1\over12}\right]
   -2{\varepsilon_-\over\varepsilon_+}
   \left[P_2\left({\mu_{0s}^0\over\varepsilon_-}\right)-{1\over12}\right]
   \right\},
\end{eqnarray}
where $\mu_{0s}^0=\varepsilon_{0F}-\hbar\Omega$.
Using the above expressions for $\nu_F$, we obtain:
\begin{eqnarray}
   \varepsilon_F&=&\nu_{0F}\sqrt{
   {[(\hbar\Omega_B')^2-(\alpha(B)/2)^2]^2-(\hbar\omega_c\alpha(B)/2)^2
   \over(\hbar\Omega_B')^2-(\alpha(B)/2)^2}}
   \left[1+{\nu_{1F}\over\nu_{0F}}+O(N^{-3/2})\right]
   \\
   {\cal E}_0'(B)&=&{2\over3}N\nu_{0F}
   \sqrt{
   [(\hbar\Omega_B')^2-(\alpha(B)/2)^2]^2-(\hbar\omega_c\alpha(B)/2)^2
   \over(\hbar\Omega_B')^2-(\alpha(B)/2)^2}[1+O(N^{-3/2})].
\end{eqnarray}

Following the similar procedure as for the zero magnetic field
(see Appendix \ref{appB}) we shall now estimate spin-orbit coupling
constant, which is now a function of the field.
Analogously to Eq.(\ref{averso}) we have (neglecting here the
spin-orbit energy):
\begin{equation}
   <\delta U>={1\over2N}\hbar\Omega_B'\cdot{\Omega_B'\over\Omega}
   \sum_{nm\sigma}(n+m+1)\cdot n_s(n\hbar\Omega+{1\over2}m\hbar\omega_c),
\end{equation}
and further:
\begin{equation}
   \alpha(B)={1\over3}\beta f_B\hbar\Omega_B'\sqrt{N},
\label{alp}
\end{equation}
with the renormalizing function:
\begin{equation}
   f_B=\left.\sqrt{1+z^2/N}\left(1-{z\over\sqrt{1+z^2}}\right)\right|
   _{z=\omega_c/2\Omega_B'}.
\label{deffB}
\end{equation}
At low fields the function $f_B$ tends to unity, while for
$B\rightarrow\infty$ it decays to zero: $f_B\sim1/\sqrt{N}$.
Therefore, to a good approximation, in the definition of $f_B$ we can
replace $\Omega_B'$ by $\Omega_0$, defined by Eq.(\ref{Om0}).

Finally, using the above formula for $\alpha(B)$, and introducing
the following function $u_B$:
\begin{equation}
   u_B=\left.
   1+\left({\omega_c/2\Omega_0}\right)^2
   {1\over1-z}\left({1\over N}-4{z\over1-z)}\right)\right|
   _{z=\beta^2f_B^2N/36}.
\label{defuB}
\end{equation}
the Hartree energy of the system can be written in the form of
Eq.(\ref{heiamf}).



\begin{figure}
\caption{
  The average ground state energy per electron as a function of
  the number of electrons in the dot.
  The classical result (stars) is taken from
  Ref.\protect\onlinecite{shikin},
  the experimental data (stars) --
  from Ref.\protect\onlinecite{demel},
  and the two curves in between (circles) are obtained within our
  model for two values of the spin-orbit coupling constant $\beta$
  (GaAs, $\hbar\omega_0=5.4$ meV).}
\label{fig1}
\end{figure}

\begin{figure}
\caption{
  The dot radius as a function of the magnetic field
  for $N=30$ electrons.
  The three curves correspond to the spin-orbit coupling constant
  $\beta$ equal to 0.0, 0.3 and 0.5 (GaAs, $\hbar\omega_0=5.4$ meV).}
\label{fig2}
\end{figure}

\begin{figure}
\caption{
  The average ground state energy per electron as a function of
  the magnetic field and the number of electrons.
  The three frames correspond to the spin-orbit coupling constant
  $\beta$ equal to 0.0, 0.3 and 0.5 (GaAs, $\hbar\omega_0=5.4$ meV).
  Insets show the chemical potentials.}
\label{fig3}
\end{figure}

\begin{figure}
\caption{
  FIR absorption spectra of a quantum dot containing 25 electrons.
  Stars -- experiment by Demel {\sl et al.} \protect\cite{demel},
  lines -- the model (GaAs, $\beta=0.3$, $\hbar\omega_0=7.5$ meV).}
\label{fig4}
\end{figure}

\end{document}